\documentclass[12pt]{article}

\usepackage{amsmath,amssymb,amsthm,amscd,pstricks}
\usepackage{graphicx,tocloft}
\usepackage[nosort]{cite}
\def\figurename{Figure}

\makeatletter

\renewcommand{\fnum@figure}[1]{\textbf{\figurename~\thefigure}:}

\makeatother

\makeatletter

\renewcommand\section{\@startsection{section}{1}{\z@}
                                   {-3.5ex \@plus -1ex \@minus -.2ex}
                                   {2.3ex \@plus .2ex}
                                   {\normalfont\large\bfseries}}
\renewcommand\subsection{\@startsection{subsection}{2}{\z@} 
                                   {-3.25ex\@plus -1ex \@minus -.2ex}
                                   {1.5ex \@plus .2ex}
                                   {\normalfont\normalsize\bfseries}}
\renewcommand\subsubsection{\@startsection{subsubsection}{3}{\z@}
                                   {-3.25ex\@plus -1ex \@minus -.2ex}
                                   {1.5ex \@plus .2ex}
                                   {\normalfont\normalsize\bfseries}}
\renewcommand\paragraph{\@startsection{paragraph}{4}{\z@}
                                   {3.25ex \@plus1ex \@minus.2ex}
                                   {-1em}
                                   {\normalfont\normalsize\bfseries}}

\makeatother

\newdimen\tableauside\tableauside=1.0ex
\newdimen\tableaurule\tableaurule=0.4pt
\newdimen\tableaustep
\def\phantomhrule#1{\haox{\vbox to0pt{\hrule height\tableaurule
width#1\vss}}}
\def\phantomvrule#1{\vbox{\haox to0pt{\vrule width\tableaurule
height#1\hss}}}
\def\sqr{\vbox{%
  \phantomhrule\tableaustep

\haox{\phantomvrule\tableaustep\kern\tableaustep\phantomvrule\tableaustep}%
  \haox{\vbox{\phantomhrule\tableauside}\kern-\tableaurule}}}
\def\squares#1{\haox{\count0=#1\noindent\loop\sqr
  \advance\count0 by-1 \ifnum\count0>0\repeat}}
\def\tableau#1{\vcenter{\offinterlineskip
  \tableaustep=\tableauside\advance\tableaustep by-\tableaurule
  \kern\normallineskip\haox
    {\kern\normallineskip\vbox
      {\gettableau#1 0 }%
     \kern\normallineskip\kern\tableaurule}%
  \kern\normallineskip\kern\tableaurule}}
\def\gettableau#1 {\ifnum#1=0\let\next=\null\else
  \squares{#1}\let\next=\gettableau\fi\next}

\newcommand{\be}{\begin{equation}}
\newcommand{\ee}{\end{equation}}
\newcommand{\bea}{\begin{eqnarray}}
\newcommand{\eea}{\end{eqnarray}}
\newcommand{\ba}{\begin{array}}
\newcommand{\ea}{\end{array}}
\newcommand{\id}{\haox{1\kern-.27em l}}

\newcommand{\ZZ}{\mathbb{Z}}
\newcommand{\CC}{\mathbb{C}}
\newcommand{\RR}{\mathbb{R}}
\newcommand{\PP}{\mathbb{P}}
\newcommand{\half}{ {\textstyle \frac{1}{2}  } }
\newcommand{\al}{\alpha}

\newcommand{\Ga}{\Gamma}

\newcommand{\ka}{\kappa}

\newcommand{\de}{\delta}

\newcommand{\ep}{\epsilon}
\newcommand{\vep}{\varepsilon}

\newcommand{\la}{\lambda}

\newcommand{\De}{\Delta}

\newcommand{\Ups}{\Upsilon}
\newcommand{\cN}{\mathcal{N}}
\newcommand{\cA}{\mathcal{A}}
\newcommand{\cO}{\mathcal{O}}

\newcommand{\cW}{\mathcal{W}}

\newcommand{\cF}{\mathcal{F}}

\newcommand{\cC}{{\mathcal C}}

\newcommand{\pa}{\partial}
\newcommand{\rar}{\rightarrow}

\newcommand{\non}{\nonumber}
\newcommand{\lb}{\langle}
\newcommand{\rb}{\rangle}

\newcommand{\SU}{\mathrm{SU}}

\newcommand{\su}{\mathrm{su}}

\newcommand{\sll}{\mathrm{sl}}
\newcommand{\U}{\mathrm{U}}
\newcommand{\ul}{\mathrm{u}}

\newcommand{\ts}{\textstyle}

\begin{document}

\begin{center}
\vspace*{-2mm}
{\Large\sf
{Coset conformal blocks and {\large $\cN\!=\!2$}  gauge theories }}

\vspace*{6mm}
{\large Niclas Wyllard}

\vspace*{4mm}

{\tt n.wyllard@gmail.com}

\vspace*{7mm}
{\bf Abstract} 
\end{center}
\vspace*{0mm}
\noindent  
It was recently suggested that  the $\widehat{\su}(N)_\ka\!  \oplus \! \widehat{\su}(N)_p / \widehat{\su}(N)_{\ka+p}$ coset conformal field theories should be related to $\cN\!=\!2$ $\SU(N)$ gauge theories on $\RR^4/\ZZ_p$.   In this paper we study various aspects of this proposal.  We perform explicit checks of  the  relation for  $(N,p)\!=\!(2,4)$, where the symmetry algebra of the coset is the so called $S_3$ parafermion algebra.  Even though the symmetry algebra of the coset is unknown for  generic $(N,p)$ models, we manage to perform non-trivial checks  in the general case by using knowledge of the Kac determinant of the coset CFT. We also find evidence that the conformal blocks  of the $(N,p$) model should factorise into a certain product of $p$ $(N,1)$ conformal blocks. Precisely this structure is present in the instanton partition function on $\RR^4/\ZZ_p$.

\vspace{1mm}

\setcounter{tocdepth}{1}

\setcounter{equation}{0}
\section{Introduction}\label{sint}

In the past two years several precise relations between $2d$ conformal field theories and $4d$ $\cN\!=\!2$ gauge theories have been discovered. One example \cite{Alday:2009a,Wyllard:2009} is the by now well-known AGT relation that connects $\SU(N)$ gauge theories on $\RR^4$ with the Toda field theories and their associated $\cW_N$ symmetry. More recently, a relation that connects  the $\SU(2)$ gauge theories on $\RR^4/\ZZ_2$ with the $\cN\!=\!1$ super-Liouville theory and its  $\cN\!=\!1$ superconformal symmetry  was proposed  \cite{Feigin:2011}. This relation was further studied in \cite{Bonelli:2011a,Belavin:2011,Bonelli:2011b}. 

These relations are special cases of a more general proposal  \cite{Nishioka:2011} that connects  $\SU(N)$ gauge theories on $\RR^4/\ZZ_p$ with the coset conformal field theories based on the coset 
\be \label{suncos}
\frac{ \widehat{\su}(N)_\ka \oplus \widehat{\su}(N)_p }{ \widehat{\su}(N)_{\ka+p} } .
\ee
Here (and throughout this paper) $p$ is a positive integer and $\ka$ is a free parameter.  Coset conformal field theories \cite{Goddard:1984} have been intensely studied in the literature\footnote{There is also a lagrangian version of the construction involving gauged WZNW models \cite{Gawedzki:1988a}.} mostly in the context of rational conformal field theories, although here we are interested in non-rational models. The AGT relation corresponds to the special case $p\!=\!1$ (it is known that when $p\!=\!1$ the extended chiral symmetry algebra of the coset (\ref{suncos}) is the $\cW_N$ algebra, see e.g.~\cite{Bouwknegt:1992}). Similarly, when $(N,p)=(2,2)$ the symmetry algebra of the coset is the  $\cN\!=\!1$  superconformal algebra and one recovers the relation in \cite{Feigin:2011}. 
Based on these two special cases the extension to the general $(N,p)$ models is very natural. Since the $p\!=\!1$ case corresponds to the Toda theories the general cases have been called (conformal) para-Toda theories, see e.g.~\cite{Nemeschansky:1991} and references therein. The $(2,p)$ models correspond to the para-Liouville theories  recently studied in \cite{Bershtein:2010}.

In \cite{Nishioka:2011} a first check of the proposed relation in the general case was performed. This check is based on a computation of the central charge of the $2d$ CFT (\ref{suncos}) from M-theory considerations, generalising the earlier  $p\!=\!1$ analysis \cite{Benini:2009} (which in turn was based on an observation in \cite{Bonelli:2009}).  Let us recapitulate the main points of this analysis.  

The central charge of the coset (\ref{suncos}) takes the well-known form 
\be \label{ccos}
c_{\rm coset}= c_{ \widehat{\su}(N)_\ka }+  c_{ \widehat{\su}(N)_p } -  c_{ \widehat{\su}(N)_{\ka+p} } = \frac{ \ka \, p \, (N^2{-}1)(\ka {+} p {+} 2 N)}{(\ka{+} N) (p {+}N) (\ka{+} p {+} N)}\,,
\ee
where we used $c_{ \widehat{\su}(N)_\ka }= \frac{\ka (N^2-1)}{\ka + N}$.  In contrast, the expression obtained from M-theory, or more precisely from the anomaly polynomial of the $6d$ $A_{N-1}$ $(2,0)$ theory, is \cite{Nishioka:2011} 
\be \label{cM}
c_{\rm M-theory} = \frac{N(p^2{-}1)}{p\!+\!N}-(p\!-\!1)+ \left[ \frac{p(N^2{-}1)}{p\!+\!N}+\frac{N(N^2{-}1)}{p}\left(\sqrt{\frac{\ep_1}{\ep_2}} +\sqrt{\frac{\ep_2}{\ep_1}} \right)^2 \right].
\ee
where $\ep_{1,2}$ are the gauge theory deformation parameters introduced in \cite{Moore:1997,Nekrasov:2002}.  The relation between $\ka$ and $\ep_{1,2}$ is 
\be \label{kaep}
\ka{+}N  = -p\, \frac{\ep_2}{\ep_1+\ep_2} = -p \frac{b^2}{1+b^2} \,,
\ee
where we also introduced the CFT parameter $b$ via the usual relation  \cite{Alday:2009a} $b^2 = \ep_2/\ep_1$.  For $p\!=\!1$ this result can be obtained from previously known facts as follows. In addition to the coset formulation, the $\cW_N$ algebras can also be obtained from the $\widehat{\sll}_N$ current algebra at level $k$ by quantum Drinfeld-Sokolov reduction (see e.g.~\cite{Bouwknegt:1992} and references therein). The level $k$ is related to $\ep_{1,2}$  via \cite{Alday:2010}
\be \label{kep}
 k{+}N = - \frac{\ep_2}{\ep_1} = -b^2.
\ee
The parameter $\ka$ appearing in the coset (\ref{suncos}) is not precisely the same as the parameter $k$. The relation between $\ka$ and $k$ is (see e.g.~\cite[Section 7.3.3]{Bouwknegt:1992})
\be \label{kka}
\ka{+}N =    \frac{k{+}N}{1-(k{+}N)} \,.
\ee
Combining (\ref{kep}) and (\ref{kka}) leads to (\ref{kaep}) with $p\!=\!1$.  

Using the relation (\ref{kaep}) one sees that  the central charge (\ref{ccos}) agrees with the part of  (\ref{cM}) in square brackets. The remaining piece coincides with the central charge of the coset
 \be \label{paracos}
 \frac{ \widehat{\su}(p)_N  }{ \widehat{\ul}(1)^{p-1} }\, .
 \ee
Therefore the precise statement is that the $\SU(N)$ gauge theory on $\RR/\ZZ_p$  should be related to coset theory (\ref{suncos}) plus the coset theory (\ref{paracos}) \cite{Nishioka:2011}. In addition, just as in the AGT relation, additional $\ul(1)$ factors may be important in certain computations.  Although the matching of the central charges says little about the details of the relation it nevertheless clearly shows the power of the intuition derived from the M-theory setup. 

The only cases where the relation between  the coset theory (\ref{suncos})  and the $\cN\!=\!2$ $\SU(N)$ gauge theory on $\RR^4/\ZZ_p$  have hitherto been checked in detail are $(N,p)=(N,1)$ and $(N,p)=(2,2)$.  In this paper we study  the relation between the coset theory and the gauge theory for general $(N,p)$ and in particular  provide some new checks.
   
For the special case $N\!=\!2$, $p\!=\!4$ it is known (see e.g.~\cite{Argyres:1991}) that the symmetry algebra of the coset  (\ref{suncos}) is the so called $S_3$ parafermion algebra \cite{Fateev:1985b}. In section \ref{sS3} we find  compelling evidence that suitably defined (irregular) conformal blocks in this theory indeed agree with the instanton partition function of the pure $\cN\!=\!2$ $\SU(2)$ gauge theory on $\RR^4/\ZZ_4$.
 
For general $(N,p)$ models the situation is much more difficult  since in the generic case not much is known about the symmetry algebra of the coset (\ref{suncos}).
In view of this an alternative more indirect approach would be useful. As we will see, the coset formulation provides some methods that allow us to perform non-trivial checks of the relation to gauge theory without actually knowing the precise form of the symmetry algebra.  
In particular, in section \ref{sGen} we make the observation that when the terms at a given order in the instanton partition function of the $\cN\!=\!2$ gauge theory are written as a single fraction, the denominator is (a simplified version of) the Kac determinant of the $2d$ CFT. Using knowledge about the Kac determinant for the general coset theory (\ref{suncos}) we are able to perform non-trivial checks  of the proposed relation. These checks are detailed enough to fix the map between the variables in the two theories.   
 
 There have been various results in the literature relating conformal blocks for different coset theories.  In particular, in \cite{Crnkovic:1989a,Lashkevich:1993} it was argued that conformal blocks for the rational version of the coset (\ref{suncos}) factorise into $p$ conformal blocks for the $p\!=\!1$ theory. In section \ref{sFac} we argue that the extrapolation of this argument to the non-rational setting should explain the known factorisation property of the instanton partition functions  on $\RR^4/\ZZ_p$ \cite{Sasaki:2006,Gasparim:2008,Bruzzo:2009} and provide the answer to a question posed  in \cite{Bonelli:2011a,Bonelli:2011b}.

In the next section we review some facts about the instanton partition functions of the $\cN\!=\!2$ $\SU(N)$ gauge theories on $\RR^4/\ZZ_p$ that will be needed in later sections. Then in section \ref{sS3} we perform explicit computations using the $S_3$ parafermion algebra and match the results to the  $\cN\!=\!2$ $\SU(2)$ gauge theory on $\RR^4/\ZZ_4$. Our checks for general models are discussed in section \ref{sGen}. In section \ref{sFac} we discuss the factorisation of coset conformal blocks and the relation to the gauge theory results. We conclude with a discussion of some open problems.

\setcounter{equation}{0}
\section{Instanton counting for  $\cN\!=\!2$ $\SU(N)$ gauge theory  on $\RR^4/\ZZ_p$   }\label{sInst}

In this section we briefly describe how one computes the instanton partition function for $\cN\!=\!2$ $\SU(N)$ gauge theory on (a smooth resolution of) $\RR^4/\ZZ_p$. These results were first obtained in \cite{Fucito:2004b}  (see also~\cite{Fujii:2005}) using the ADHM construction given in \cite{Kronheimer:1990}. Some applications and further developments were discussed in \cite{Fucito:2006,Dijkgraaf:2007}.  For simplicity we focus on the case of the pure $\cN\!=\!2$ $\SU(N)$ gauge theory. 

The instanton partition function for the  $\cN\!=\!2$ $\SU(N)$ theory on $\RR^4$ (i.e.~without orbifolding) is given by the well-known expresssion 
\be \label{Zp=1}
Z_N(\vec{a}, \ep_1,\ep_2,y) = \sum_{\vec{Y}} \prod_{i,j=1}^N \prod_{s\in Y_i}  \frac{1}{E_{i,j}(s,\ep_1,\ep_2)[ \ep_1{+}\ep_2-E_{i,j}(s,\ep_1,\ep_2)  ]}\,  y^{k} \,,
\ee
where the sum is over all vectors of $N$ Young tableaux $\vec{Y} = (Y_1,\ldots,Y_N)$, $k$ is the total number of boxes in $\vec{Y}$, $s$ denotes a box in the Young diagram $Y_i$, and 
\be
E_{i,j}(s,\ep_1,\ep_2) = a_i-a_j  - L_j(s) \,  \ep_1 + (A_i(s){+}1)\,  \ep_2 \,,
\ee
where $A_j(s)$ ($ L_j(s)$) is the arm (leg) length, i.e.~the number of boxes in $Y_j$ above (to the right of) the box $s\in Y_i$.  
For further details about the notation we refer to the original papers~\cite{Nekrasov:2002,Flume:2002}.

To treat the case with orbifolding one needs to take into account the action of $\ZZ_p$ on the instanton moduli space and decompose the various quantities into representations of $\ZZ_p$.  The  representations of the abelian group $\ZZ_p$ are one-dimensional and are labelled by an integer $\ell=0,1,\ldots,p-1$.  The building blocks of the instanton partition function are similar to the case without orbifolding but with some differences. The sum over fixed points still involves $N$ Young tableaux. The analogues of $N$ and $k$ are $p$-dimensional vectors with components $N_\ell$ and $k_\ell$, and to each Young diagram $Y_i$ one associates a $\ZZ_p$ charge (representation)  $r_i\in 0,\ldots,p-1$. The number of Young tableaux associated with the $\ell$th representation is equal to $N_\ell$ and $\sum_\ell N_\ell = N$. Since $k$ is replaced by a vector one has $p$ ``instanton numbers". These are conveniently organised into $k$ and $u_\ell$ with $\ell =1,\ldots,p-1$, defined as follows 
\be
u_\ell = N_\ell + k_{\ell+1}+k_{\ell-1}-2k_{\ell}  \,, \qquad \qquad  k =   \sum_\ell k_{\ell} +  \half \sum_\ell  ( p{-}\ell ) \, \ell  \, u_{\ell} \,,
\ee
where in the definition of $u_\ell$ the periodic identification $k_{\ell+p} = k_\ell$ is understood. The associated expansion parameters will be denoted $x_\ell$ and $y$. The $x_\ell$ variables are  related to the non-trivial two-cycles (exceptional divisors) in the geometry.  This situation is analogous to the case with surface operators, see e.g.~\cite{Alday:2010}. Let us also mention that it is more common in the literature to define $k$ with an extra factor of $1/p$. Such a definition also corresponds more closely to the CFT formulation as we will see later. Nevertheless, we keep the above definition since it simplifies some formul\ae{}.

With the above definitions the instanton partition function for the  $\cN\!=\!2$ $\SU(N)$ gauge theory on $\RR^4/\ZZ_p$ can be written \cite{Fucito:2004b}
\be \label{ZNp}
Z_N^{(p)} = \sum_{\{\vec{Y},\vec{r}\}} \prod_{i,j,s\in \Diamond}  \frac{1}{E_{i,j}(s,\ep_1,\ep_2)[ \ep_1{+}\ep_2-E_{i,j}(s,\ep_1,\ep_2)  ]} \, y^{k} \, \prod_{\ell=1}^{p-1} x^{u_\ell}_\ell\,.
\ee
Here the product over $i,j=1,\ldots,N$ and $s\in Y_i$ is restricted to the set $\diamond$ that comprises all $i,j,s$ that satisfy
\be
A_i(s) + L_j(s)  + 1 - r_i + r_j = 0 \!\!\! \mod p\,.
\ee

Our main interest in this paper is the case for which $u_\ell=0$ for all $\ell$. This restriction gives strong constraints and leads to gaps in the instanton expansion, i.e.~not all powers of $y$ appear in the expansion.  This is a generic feature for general $p$. 

Let us discuss how this comes about in a bit more detail.  First we write $k_\ell = k_0 + t_\ell$ for $\ell=1,\ldots,p{-}1$  which implies that $N_\ell - C_{\ell \tilde{\ell}}\, t^{\tilde{\ell}}=u_\ell = 0$ where  $C_{\ell \tilde{\ell}}$ is the $(p{-}1){\times}(p{-}1)$ Cartan matrix of the $A_{p-1}$ Lie algebra. This implies that $t^\ell= C^{\ell \tilde{\ell}}\, N_{\tilde{\ell}}$ where the inverse of the Cartan matrix is 
\be
C^{\ell \tilde{\ell}} = \min(\ell, \tilde{\ell}) -  \frac{\ell \, \tilde{\ell}}{p}  \,.
\ee
It is clear from their definition that the $t^\ell$ are integers as are the $N_\ell$, but  in general $t^\ell= C^{\ell \tilde{\ell}}\, N_{\tilde{\ell}}$ does not map integers to integers hence not all configuration are allowed. 

As an example, consider the case $N\!=\!2$ where either one $N_\ell$ is 2 or two are 1 with the rest 0. Looking at the form of the inverse Cartan matrix it is easy to see that the only possibilities that give integer $t^\ell$'s are: $N_0=2$; $N_\ell=N_{p-\ell} =1$ for $\ell=1,\ldots,\lfloor \frac{p}{2} \rfloor$; and $N_{p/2}=2$ (the latter case appears only if $p$ is even). More explicitly we have for  $(N,p)=(2,4)$
\be
(2,0,0,0)\,, \;\;  k=4m \,; \qquad (0,1,0,1)\,,   \;\;  k=4m{+}3 \,; \qquad (0,0,2,0)\,, \;\; k=4m{+}4 \,,
\ee
where $m=k_0$ is a non-negative integer. Another example is $(N,p)=(2,7)$
\bea \label{N27}
&& (2,0,0,0,0,0,0)\,, \quad k=7m \,; \quad \; \, \, \, \qquad (0,1,0,0,0,0,1)\,,  \quad k=7m{+}6 \,;\non \\
 && (0,0,1,0,0,1,0)\,,  \quad k=7m{+}10 \,; \qquad (0,0,0,1,1,0,0)\,,  \quad k=7m{+}12 \,.
\eea
Our final example is $(N,p)=(3,4)$
\bea
&&  (3, 0, 0, 0) \,,\quad  k=4m   \,; \qquad \qquad (1, 1, 0, 1)  \,,\qquad  k=4m{+}3  \,;  \non \\
&&   (1, 0, 2, 0) \,,  \quad  k=4m{+} 4   \,; \qquad \; \! \left.\ba{c} (0, 0, 1, 2) \\ (0, 2, 1, 0)\ea \!\! \right\} \,, \quad k=4m{+}5 \,.
\eea

Before we give some explicit examples of instanton partition functions on  $\RR^4/\ZZ_p$ let us mention that there exists an alternative way to organise the instanton expansion, first discussed in the simplest case $\RR^4/\ZZ_2$ in \cite{Sasaki:2006}. The formalism for a general noncompact toric surface (of which $\RR^4/\ZZ_p$ is an example)  was discussed in \cite{Gasparim:2008,Bruzzo:2009}. In this approach  the fixed point set is labelled by $p\cdot N$ Young tableaux together with $(p{-}1)\cdot N$ integers. This approach will be further discussed in section \ref{sFac}.

We close this section with some explicit examples. As above $Z_N^{(p)}$ denotes the partition function for pure $\SU(N)$ on $\RR^4/\ZZ_p$.  For the $\SU(2)$ theory on $\RR^4/\ZZ_4$ the first couple of terms are
\bea \label{Z24}
\!\! Z_2^{(4)} \!\! \!\! &= &  1   -\frac{8 (4 a^2 - 3 \ep_1^2 - 3\ep_2^2 - 10 \ep_1 \ep_2 )}{A_{1,3}\, A_{3,1}}  y^3  - \frac{2(a^2 - \ep_1^2 - \ep_2^2 - 5 \ep_1 \ep_2 )}{  \ep_1 \, \ep_2 \, A_{1,1} \, A_{2,2}  }     y^4  
\\& +&\! \!  \!   \frac{   4 (16 a^4 {-} 208 a^2 (\ep_1^2{+}\ep_2^2){-} 288 a^2 \ep_1 \ep_2{+} 147 (\ep_1^4{+}\ep_2^4)  {+} 1540 \ep_1 \ep_2 (\ep_1^2{+}\ep_2^2)  {+} 3794 \ep_1^2 \ep_2^2  )}{\ep_1 \, \ep_2\,  A_{1,3} \, A_{3,1} \, A_{1,7}\, A_{7,1}  } y^7 \non 
 \\  &  +&  \! \! \!   \frac{  8 a^4 {-} 58 a^2(\ep_1^2{+}\ep_2^2){-} 120 a^2 \ep_1 \ep_2  {+} 50 (\ep_1^4{+}\ep_2^4)  {+} 645 \ep_1 \ep_2(\ep_1^2{+}\ep_2^2)  {+} 1526 \ep_1^2 \ep_2^2}{4  \ep_1^2 \, \ep_2^2 \, A_{1,1} \, A_{2,2} \,  A_{1,5} \, A_{5,1}} y^8 + \ldots \non
\eea
where $a_1=-a_2=a$ and we used the short-hand notation 
\be \label{Ars}
A_{r,s} = (2a - r\, \ep_1 - s\, \ep_2)(2a + r\, \ep_1 + s\, \ep_2)\,.
\ee
For the $(N,p)=(2,7)$ theory we have
\bea \label{Z27}
Z_2^{(7)} \!\!\!&=& \!\!\! 1 - \frac{ 14 (4 a^2 - 6 \ep_1^2 - 6 \ep_2^2- 37 \ep_1 \ep_2 )}{A_{1,6} \, A_{6,1} }\,  y^6 -\frac{2}{7 \, \ep_1 \, \ep_2\, A_{1,1} } \, y^7
 \\ &+& \!\!\!\!
 \frac{ 42 (16 a^4 {-} 20 a^2( \ep_1^2{+} \ep_2^2) {-} 256 a^2 \ep_1 \ep_2  {+} 40 (\ep_1^4 {+}\ep_2^4) {+} 286 \ep_1 \ep_2( \ep_1^2{+} \ep_2^2)  {+} 573 \ep_1^2 \ep_2^2  )}{A_{1,4} \, A_{4,1} \, A_{2,5} \, A_{5,2} } y^{10} + \ldots \non 
\eea
Finally for the $(N,p)=(3,4)$ theory, i.e.~the $\SU(3)$ theory on $\RR^4/\ZZ_4$, we have
\be
Z_3^{(4)} = 1 + \frac{P_{10} (a,\ep_1,\ep_2)}{\prod_{i>j} \cA_{1,3}(i,j)\cA_{3,1}(i,j)}  y^3 + \frac{P_{8}(a,\ep_1,\ep_2)}{ \ep_1 \ep_2  \prod_{i>j} \cA_{1,1}(i,j)  \cA_{2,2}(i,j) }  y^4 + \ldots
\ee
where we used the short-hand notation 
\be
\cA_{r,s}(i,j) = (a_i-a_j - r\, \ep_1 - s\, \ep_2)(a_i-a_j + r\, \ep_1 + s\, \ep_2)\,,
\ee
and $P_d(a,\ep_1,\ep_2)$ is a homogeneous polynomial in $a_i$ and $\ep_{1,2}$ of degree $d$. These polynomials are known but somewhat complicated.

\setcounter{equation}{0}
\section{$S_3$ parafermions and $\cN\!=\!2$ $\SU(2)$ gauge theory on $\RR^4/\ZZ_4$ }\label{sS3}

For the case $N=2$, $p=4$ it is known (see e.g.~\cite{Argyres:1991}) that the extended symmetry algebra of the coset (\ref{suncos}) is  isomorphic to the so called spin $4/3$ $S_3$ parafermion algebra \cite{Fateev:1985b}.

We start by  recapitulating the main facts about this algebra. We will be brief since excellent discussions can be found in \cite{Fateev:1985b} and section 2 of  \cite{Argyres:1993}. In addition to the stress tensor $T(z)$  the algebra also contains two spin $4/3$ fields $G^{\pm}(z)$. The operator product expansions are
\bea \label{S3OPE}
T(z)T(w)&\sim& \frac{1}{(z-w)^4}\bigg[  \frac{c}{2} +  2 (z-w)^2 T(w) + (z-w)^3 \pa T(w) \bigg], \non \\
T(z)G^{\pm}(w)&\sim& \frac{1}{(z-w)^2}\bigg[  \frac{4}{3} G^{\pm}(w) + (z-w) \pa G^{\pm} (w) \bigg] ,\non \\
G^{\pm}(z)G^{\pm}(w)&\sim& \frac{\la^{\pm} }{(z-w)^{4/3} }\bigg[  G^{\mp}(w) + \frac{1}{2} (z-w) \pa G^{\mp} (w) \bigg], \\
G^{+}(z)G^{-}(w)&\sim& \frac{1 }{(z-w)^{8/3} }\bigg[  \frac{3c}{8} +  (z-w)^2 T(w) + (z-w)  \bigg] ,\non 
\eea
where
\be
c = 2 +\frac{3}{2}\left(b+\frac{1}{b}\right)^2 \,, \qquad \la^{\pm} = \pm \sqrt{\frac{c-8}{6}} \,.
\ee
One can associate a $\ZZ_3$ charge $q$ to the fields of the algebra such that $G^{\pm}$ has charge $q=\pm 1$ and $T$ has charge $q=0$. This charge assignment is consistent with the above OPEs; note that e.g.~$G^- G^-$ has $\ZZ_3$ charge $q=+1$.

The first two OPEs in (\ref{S3OPE}) can easily be translated into commutation relations for the associated modes
\bea
[L_m,L_n] &=& (m-n) L_{n+m} + {\ts \frac{c}{12}} (m^3-m) \de_{n+m,0} \,, \non \\
{}[L_m,G^{\pm}_r] &=& ({\ts \frac{m}{3}} -r)G^{\pm}_{m+r}\,,
\eea
but the fractional powers in the last two OPEs in (\ref{S3OPE}) complicates the analysis and requires a careful treatment.  

In general for the $G^{\pm}_r$ modes one has $r\in \ZZ/3$, but it turns out that not all modes can act on a state with a given $\ZZ_3$ charge,  so it is very important to keep track of the $\ZZ_3$ charge of the state upon which the  $G_r^{\pm}$ modes act. If we denote a generic state of charge $q$ by $| q\rb$ then the allowed combinations are
\be \label{S3G}
 G^\pm_{m-(1\mp q)/3} |q\rb \,, \qquad m\in \ZZ\,.
\ee
Similarly, the commutation relations of the $G^{\pm}$ modes also depend on the charge of the state upon which they act and take the following rather complicated forms (known as generalised commutation relations)
\bea \label{gencomm}
&& \sum_{k=0}^{\infty}  C^{(-2/3)}_{k} \left[ G^{\pm}_{ \pm \frac{q}{3}+n-k } G^{\pm}_{ \frac{2\pm q}{3}+m+k } {-} G^{\pm}_{ \pm \frac{q}{3}+m-k } G^{\pm}_{ \frac{2\pm q}{3}+n+k }   \right] \! |q\rb = \frac{ \la^{\pm} }{2}(n{-}m)G^{\mp}_{ \frac{2\pm 2q}{3}+n+m }   |q\rb \non \\
&&\sum_{k=0}^{\infty}   C^{(-1/3)}_{k} \left[ G^{+}_{\frac{1+q}{3}+n-k } G^{-}_{ -\frac{1+ q}{3}+m+k } {+} G^{-}_{ -\frac{2+q}{3}+m-k } G^{+}_{ \frac{2+ q}{3}+n+k }   \right] \! |q\rb  \\ 
&& \qquad \qquad \! = \left[ L_{n+m} + {\ts \frac{3c}{16} } (n+1+{\ts \frac{q}{3} }) (n+{\ts \frac{q}{3} }) \right] \! |q\rb \non \,,
\eea
where
\be
C^{(\nu)}_k =  (-1)^k \binom{\nu}{k}= \frac{(-1)^k}{k!} \prod_{i=1}^k (\nu-i+1) \,. 
\ee
Highest weight states of the above algebra are as usual annihilated by the positive modes, i.e.~by $L_m$ and $G^{\pm}_r$ with $m,r>0$. There are three different classes of highest weight states conventionally denoted  S, D and R. The S and D cases are analogous to the Neveu-Schwarz sector of the superconformal algebra whereas R is analogous to Ramond sector. Here we will only consider the S and D cases. A highest weight state in the S sector has $q=0$ whereas a highest weight state in the D sector has $q=-1$ (the choice $q=+1$ describes an isomorphic module).  We denote these highest weight states $|\al;q\rb$. The conformal dimensions ($L_0$ eigenvalues) are 
\be
\De_{\al}^{(S)} ={\ts  \frac{1}{4}} \al\, (b+{\ts \frac{1}{b}} -\al) \,, \qquad
\De_{\al}^{(D)} = {\ts  \frac{1}{4}} \al\, (b+{\ts \frac{1}{b}} -\al)+ {\ts \frac{1}{12}}\,.
\ee
The  descendants of a highest weight state take the usual form 
\be
G^{\pm}_{-r_1}\cdots G^{\pm}_{-r_{u}} L_{-n_1}\cdots L_{-n_{v}} |\al;q\rb \,.
\ee
We define the level of a descendant as $n= n_0 + \de$, where $n_0 = \sum_{i=1}^u r_i +  \sum_{j=1}^v n_j  $, and $\de=0$ in an S module and $\de=\frac{1}{12}$ in a D module.
From (\ref{S3G}) one deduces that the levels of the descendants in an S module belong to $n\in \ZZ $ or $n\in \ZZ {+} \frac{1}{3}$  whereas in a D module one finds $n\in \ZZ {+}\frac{3}{4}$ or $n\in \ZZ{+} \frac{1}{12}$.

Even though the commutation relations of the $G^\pm$ modes are somewhat complicated, most of the usual operations can still be performed. It can be shown that $G^\pm_{r}$ annihilate a descendant provided that $r$ is greater than the value of $n_0$ for that state. Moving a positive $G^\pm$ mode to the right through the string of negative modes of a descendant is accomplished using the above generalised commutation relations (\ref{gencomm}). Even though infinite sums appear in these relations, only a finite number of terms actually contribute since for $k$ large enough the terms will annihilate the state upon which they act,  so the infinite range of the sums is not a problem for practical computations.
Our conventions for Hermitian conjugation are: $(L_n)^{\dag} = L_{-n}$ and $(G^\pm_r)^{\dag}=-G_{-r}^{\mp}$. 

Having reviewed the facts that we need, we turn to the main point of this section. According to the proposal discussed in the introduction, conformal blocks for the $S_3$ parafermion algebra discussed above should be related to instanton partition functions for the $\cN\!=\!2$ $\SU(2)$ gauge theory formulated on $\RR^4/\ZZ_4$. For simplicity we focus on the instanton partition function for the pure $\SU(2)$ theory without matter that is expected to be dual to irregular conformal blocks that arise from the norm of a certain Whittaker vector.    

Let us briefly recall how the analogous results worked for the $(N,p)=(2,1)$ case (Virasoro algebra \cite{Gaiotto:2009b,Marshakov:2009}\footnote{For this case the relation was proven in \cite{Hadasz:2010}, see also \cite{Fateev:2009}.}) and the $(N,p)=(2,2)$ case (NS sector of $\cN\!=\!1$ superconformal algebra \cite{Feigin:2011}). There is a state $|W\rb$ (the Whittaker vector) whose norm can be computed perturbatively by inserting a complete set of states 
\be
\sum_{{\bf n}, {\bf n}'} \lb W| {\bf n}';\al  \rb X_{\al}^{-1}({\bf n}';{\bf n} )  \lb {\bf n}; \al |W\rb  \,.
\ee
Here $|{\bf n}; \al \rb$ denote descendants of the intermediate primary state $|\al\rb$ and $X^{-1}$ is the inverse of the Gram/Shapovalov matrix of inner products of the descendants. Because of the properties of $|W\rb$ only a subset of descendants have non-vanishing inner products with $|W\rb $ so at a given level only a particular component of the inverse of  the Gram/Shapovalov matrix gives a non-zero contribution. For the $(2,1)$ case the contributing descendants at level $n\in \ZZ^+$ are $(L_{-1})^n|\al\rb$ with $L_1 | W\rb= \sqrt{z} |W\rb$, and for the $(2,2)$ case the contributing descendants at level $n\in \ZZ^+/2$ are $(G_{-1/2})^{2n}|\al\rb$ with $G_{1/2} | W\rb = z^{1/4}  |W\rb$. The (diagonal) component of $X^{-1}$ corresponding to these descendants is equal to the instanton partition function at order $p \, n$ and $z^{1/p}$ is proportional to the instanton expansion parameter $y$.
This alternative way of presenting the relation between the CFT and gauge theory quantities (first emphasised in \cite{Marshakov:2009}) will be used in what follows. In other words, we are looking for a relation between a component of the inverse of the Gram/Shapovalov matrix  for the $S_3$ parafermion algebra at level $n$ and the instanton partition function of the $\cN\!=\!2$ $\SU(2)$  theory on $\RR^4/\ZZ_4$  at order $y^{4n}$. 

Before doing any actual computations there is a simple check that can be performed. It is clear that for the proposed relation to make sense $4n$ has to be an integer since in our conventions only integer powers appear in the instanton expansion. From the above results we see this is possible at level $m$ in the S sector and at level $m{+}\frac{3}{4}$ in the D sector. Thus we expect contributions at orders $4m$ and $4m+3$ in the instanton expansion, but we know from the results in the previous section that these are precisely the orders  at which there are non-vanishing contributions. 
Note that there are also descendants with levels such that $4n$ is not an integer ($n\!=\!m{+}\frac{1}{3}$ in the S sector and $n\!=\!m{+}\frac{1}{12}$ in the D sector). At such levels apparently no components can have gauge theory interpretations of the type we have discussed. We return to this point below but for now we focus on the levels for which $4n$ is an integer.

Thus we need to compute the inverse of the Gram/Shapovalov matrix at levels $m$ in the S sector and at levels $m{+}\frac{3}{4}$ in the D sector. Similar computations were carried out in \cite[Section 3]{Kakushadze:1993}. In that paper the number of independent descendants at a given level was also determined. 

At level 3/4 a basis of suitably normalised descendants can be chosen as 
\be
| 1 \rb = \sqrt{2}\, G_{-2/3}^+ |\al; -1\rb \,, \qquad | 2 \rb = \sqrt{2} \,G^{-}_{-2/3} G^-_{0} |\al; -1\rb \,, 
\ee
and the Gram/Shapovalov matrix becomes 
\be
2 \left( \ba{cc} \De_\al^{(D)} {+} {\ts \frac{c}{12} } &  \sqrt{\frac{c-8}{6} }( \De_\al^{(D)} {-} {\ts \frac{c}{24}} ) \\  \sqrt{\frac{c-8}{6} }( \De_\al^{(D)} {-} {\ts \frac{c}{24}} ) &( \De_\al^{(D)} {+} {\ts \frac{c}{12} } )( \De_\al^{(D)} {-} {\ts \frac{c}{24} }) \ea \right) 
\ee
The $1,1$ component of the inverse is (including the $z^{3/4}$ factor)
\be
\frac{  b \, (-1 - 2 \al \, b - 4 b^2 + 2 \al^2 \, b^2 - 2 \al\, b^3 - b^4) }{ (\al + b) (1 + 
   \al \, b) (-1 + \al\, b - 2 b^2) (-2 + \al \, b - b^2) } \, z^{3/4} \,.
 \ee
At level 1 a basis of suitably normalised descendants can be chosen as 
\be
| 1 \rb = 4 \,G^+_{-2/3}G^-_{-1/3} |\al; 0\rb \,, \qquad | 2 \rb =  4\, G^-_{-2/3}G^+_{-1/3} |\al; 0\rb \,, 
\ee
Note that $|1\rb+|2\rb = L_{-1}|\al;0\rb$. The Gram/Shapovalov matrix becomes 
\be
 \frac{4^2 \De^{(S)}_\al}{3} \left( \ba{cc} 2\De^{(S)}_\al  {+} {\ts \frac{c}{4}}{+}1 &  2\De^{(S)}_\al  {+} {\ts \frac{c}{4}}{-}2\\ 2\De^{(S)}_\al  {+} {\ts \frac{c}{4}}{-}2&  2\De^{(S)}_\al  {+} {\ts \frac{c}{4}}{+}1 \ea \right) 
\ee
The $1,1$ component of the invese is  (including the $z$ factor)
\be
-\frac{ b (-3 - 4 \al\, b - 18 b^2 + 4 \al^2 \,b^2 - 4 \al\, b^3 - 3 b^4)}{8
 \al (-3 + 2 \al\, b - 3 b^2) (-1 + \al\, b - b^2) (1 + 2 \al\, b + b^2)} \, z\,.
\ee

Comparing these expressions to the $y^3$ and $y^4$ terms in the gauge theory expression (\ref{Z24}) we find precise agreement provided we identify 
\be
a = \al - \half (b + {\ts \frac{1}{b}}) \,, \quad \ep_1 = 1/b \,, \quad \ep_2 =b \,, \quad y = - z^{1/4} \,.
\ee

\setcounter{equation}{0}
\section{Checks for general $(N,p)$ models via Kac determinant }\label{sGen}
 
In the previous section we found compelling evidence for a relation between the coset theory (\ref{suncos}) with $(N,p)=(2,4)$ and the $\cN\!=\!2$ $\SU(2)$ theory on $\RR^4/\ZZ_4$. This analysis made use of the symmetry algebra of the coset --- the $S_3$ parafermion algebra ---  which for this case is known in explicit form. Ideally one would like to extend this method to the cases with general $N$ and $p$. However, this is far from straightforward since the symmetry algebra of the coset with generic  $N$ and $p$ is not known in explicit form. Strictly speaking this is also true for the $(N,1)$ models with $N$ large since the corresponding $\cW_N$ algebras have not been explicitly written down, but  many things are known about these algebras and construction algorithms exist so this is merely a technical problem. In contrast for $p> 2$ the problem is more severe.  

To describe the problems that arise we focus on the $N\!=\!2$ models  with general $p$. For these models it is known that the symmetry algebra includes a field $G(z)$ of conformal dimension $\Ups=(p{+}4)/(p{+}2)$ \cite{Kastor:1987,Argyres:1991}. It is possible to write down the OPE for this field with itself; schematically it is given by  
\be
G(z) G(w) \sim (z-w)^{-2\Ups} (1+\ldots) + \la_p (z-w)^{-\Ups}(G(w)+\ldots)\,,
\ee
where $\la_p$ is known \cite{Argyres:1991}. The main problem is that the right-hand side contains singularities involving two different fractional powers.  Such models are called non-abelianly braided and  are difficult to analyze. In contrast, models with only a single fractional power are called abelianly braided (or parafermionic) and can be analyzed as we saw in one example in the previous section.  

 For $p\!=\!4$ the model naively is non-abelianly braided, but it  can be transformed into an abelianly braided model (the $S_3$ parafermion algebra). Generically such miracles do not occur. The fact that $p\!=\!4$ is special can be traced to the existence of the conformal embedding of $\widehat{\su}(2)_4$ inside $\widehat{\su}(3)_1$ together with the fact that abelianly braided theories are intimately connected with current algebras with level 1 (see e.g.~\cite{Bagger:1988} for a discussion). 
   
Since the symmetry algebra is not known in the general case,  we appear to be at an impasse. However, even though we do not know much about the symmetry algebra there are still some things that are known about the coset theories that can be profitably used to perform quite non-trivial checks of the proposed relation to $\cN\!=\!2$ gauge theories on $\RR^4/\ZZ_p$.  

To motivate the methods that we will use, let us revisit the $(2,4)$ model and ask how much could have been inferred without using the explicit knowledge about the symmetry algebra.  
Recall that a characteristic feature of the instanton expansion is that only certain powers of the instanton expansion parameter occur. This feature is mirrored on the CFT side by the fact that the  descendants only occur at certain levels, and  we were able prove to all orders that the same pattern appears as in the instanton expansion. For this analysis one does not need to know the commutation relations of the algebra so this simple check can be extended to more general models as we will see below. 

 In fact we can do a bit more. Recall that the terms in the instanton expansion with a given instanton number were identified with a certain component of the inverse of the Gram/Shapovalov matrix at the corresponding level. From the general formula for the inverse of a matrix (Cramer's rule)  it follows that a fixed matrix element in the inverse of the Gram matrix/Shapovalov is a rational function where the denominatior is the determinant of the Gram/Shapovalov matrix.  The determinant of the Gram/Shapovalov matrix is also known as the Kac determinant. 
  In general there may be cancellations, so that when written in factorised form the denominator only has to be contained in the Kac determinant; in other words each factor in the denominator also appears in the Kac determinant. The point is that the Kac determinant is known also for cases where the symmetry algebra is not known,  which will allow us to check if the instanton expansion has the right structure for general models.   
  
When $N\!=\!2$ the Kac determinant at level $n$ has the schematic form 
\be \label{genKac}
\prod_{\substack{r,\,s} }   (\De(\al) -\De_{r,s})^{P(n,r,s) }\,,
\ee
 where $\De(\al)$ is the conformal dimension of the highest weight state $|\al\rb$ and the zeros $\De_{r,s}$ are parametrised by positive integers $r,s$ that are subject to certain restrictions which depend on the level $n$. The $P(n,r,s)$ exponents are certain integers that determine the orders of the zeroes.  
  As mentioned above,  in the denominator of the relevant component  of the inverse of the Gram/Shapovalov matrix the full Kac determinant need not appear. Stated differently the $P(n,r,s)$ may be different in the denominator. In fact we will make the more specific conjecture that all types of $\De_{r,s}$ factors  that appear in the Kac determinant with {\it non-zero} $P(n,r,s)$ also appear in the denominator but with {\it single} powers only. If correct, this implies that when written as a single fraction the terms at a given order in the instanton expansion should have a denominator that agrees (up to a constant)  with (\ref{genKac}) with all (non-zero) $P(n,r,s)$ replaced by 1 (implying that at each order in the instanton expansion  the expression only has simple poles).
 
We now proceed to perform the checks we  just outlined in more detail. We start with the $N\!=\!2$ case with general $p$. The main reference is \cite{Kakushadze:1993}  (see also \cite{Bershtein:2010}). 

First we describe the possible types of representations/highest weight states. There are $p$ types of representations and the corresponding vertex operators will be denoted $V^{(\ell)}_\al(z)$ with $\ell=0,1,\ldots,p-1$. The associated states will be denoted $|\al;\ell\rb$.  The conformal dimensions are
\be \label{N2dim}
\De^{(\ell)}(\al) = \frac{\al(Q-\al)}{p} + \frac{\ell(p-\ell)}{2p(p+2)}\,,
\ee
where as usual $Q=b+1/b$. Note that this expression is symmetric under $\ell\rar p-\ell$ so $\ell$ and $p-\ell$ describe the ``same" representation and form a doublet. Therefore there are three classes: $\ell=0$, $\ell\oplus p\!-\!\ell$ for $\ell=1,\ldots,\lfloor \frac{p-1}{2} \rfloor$ where $\lfloor \cdot \rfloor$ denotes the integer part, and finally if $p$ is even $\ell=p/2$.  These three cases will be collectively  labelled by $\ell=0,\ldots, \lfloor \frac{p}{2} \rfloor$ from now on. The connection to the notation used in the previous section for $p\!=\!4$ is as follows: $\ell=0$ corresponds to S§, $\ell=2$ to D and $\ell=1$ to R. 

The level in the $\ell$th sector of a descendant is defined as \cite{Kakushadze:1993} 
\be 
n=n_0 + \frac{ \ell( p{-}\ell)}{2p(p+2)} \,, 
\ee
 where $n_0$ is the ``conventional"  level arising from summing up the mode numbers in the descendant (compare the discussion in the previous section). It is known that the field $G(z)$ mentioned above  generates the full symmetry algebra so to determine the possible levels it is sufficient to consider this field. The possible modes of $G(z)$ acting on a generic state, $|\ell\rb$, in the $\ell$th sector are \cite{Kakushadze:1993}  (here $m$ is a non-negative integer)  
\be
G_{-m-(\ell+2)/(p+2) }|\ell\rb= |\ell{+}2\rb\,, \quad \;\; G_{-m}|\ell\rb= |\ell\rb\,, \quad \;\;  G_{-m -(p+2- \ell)/(p+2) }|\ell\rb= |\ell{-}2\rb\,,
\ee
where the schematic notation indicates to which sector the action of the mode takes the state. There are some restrictions:  For $\ell=0$ only the first case is possible, for $\ell=1$ the last case is excluded, and when $\ell{+}2>p/2$, $|\ell{+}2 \rb$ should be identified with $|p{-}\ell {-}2\rb $. 

Recall from the discussion above that for the comparison with the gauge theory results we are interested in the levels that satisfy the constraint that $p\, n$ is an integer. As an example, let us work out the  $p\!=\!7$ case in detail. In this case we have $n_0\!=\!\frac{k}{9}$ for some $k$ (not all $k$ are possible!).  For descendants of a $\ell=0$ state we have $n\!=\!n_0$ so we need $k=9m$ to satisfy the constraint, which is possible (arising for instance from $(G_{-7/9} G_{-2/9})^m|\al;\ell=0\rb$). For $\ell=1$ we have $n=\frac{k}{9}{+}\frac{3}{7\cdot9}$. For $p\,n$ to be integer the only possibility is $k=9m+6$. This is possible but not for $m=0$ so we need to replace $m\rar m{+}1$ leading to $n=m{+}\frac{12}{7}$. For $\ell=2$ we have $n=\frac{k}{9}{+}\frac{5}{7\cdot9}$ and need $k=9m{+}7$ which is possible, giving $n=m{+}\frac{6}{7}$. Finally for $\ell=3$ we have $n=\frac{k}{9}{+}\frac{6}{7\cdot9}$ and need $k=9m{+}3$, which is possible but not for $m=0$, hence $n=m{+}\frac{10}{7}$. Summarising, the allowed values with $p\,n$ an integer are
\be \label{p7levels}
7m \quad(\ell{=}0)\,; \qquad \; 7m {+}12 \quad(\ell{=}1)\,; \qquad  \;  7m {+}6\quad(\ell{=}2)\,; \qquad   \; 7m{+}10 \quad(\ell{=}3) \,,
\ee
which precisely agree with the instanton result, cf.~(\ref{N27}).  We have checked several other values of $p$ finding agreement in each case. It may be possible to devise a general proof for arbitrary $p$ but we have not attempted to do so.

Next we turn to the analysis of the Kac determinant. When stripped of the zero-mode part 
 it takes the following form in the $\ell$th sector  at level $n$  \cite{Kakushadze:1993}
\be \label{N2Kac}
\prod_{\substack{r\,s\le p \,n \\[1pt]  |r-s| = \pm \ell \! \! \! \! \mod \! p } }  \!\!  \! \! \! \!   \! \!   (\De^{(\ell)}(\al)-\De^{(\ell)}_{r,s})^{P_{\tilde{\ell}}(n-rs/p)} \,,
\ee
where the product is over all positive integers $r,s$ subject to the given conditions and $\tilde{\ell} = r{+}s \!\! \mod p$.  The precise values of the $P_{\tilde{\ell}}(n{-}rs/p)$ are given in \cite{Kakushadze:1993} but will not be important; we only need to know that they are non-zero unless the corresponding $z^{n-rs/p}$ term in the expansion in the $\min(\tilde{\ell},p{-}\tilde{\ell})$ sector is absent meaning that there are no states at that level.
The expressions for the zeros are given by  \cite{Kastor:1987,Kakushadze:1993} 
\be \label{Ders}
\De^{(\ell)}_{r,s} = \frac{c- c_0}{24} +\frac{1}{96} \bigg[ (r+s) \sqrt{c_0-c} + (r-s)\sqrt{c_1-c}\,  \bigg]^2 + \frac{ \frac{\ell}{2}( \frac{\ell}{2}+1)}{p+2} - \frac{\ell^2}{4p}\,, 
\ee
where $c_0=3p/(p+2)$, $c_1=c_0 + 24/p$, and the central charge is, cf.~(\ref{ccos}), (\ref{kaep})
\be
c = \frac{3p}{p+2} +\frac{6}{p} \left(b+\frac{1}{b}\right)^2 .
\ee
Inserting these expressions into (\ref{Ders}) and remembering the result (\ref{N2dim}) we find that the factors in the product  (\ref{N2Kac}) can be simplified to
\be
\De^{(\ell)}(\al)-\De^{(\ell)}_{r,s} =  \frac{\al(Q-\al)}{p}  - \frac{Q^2}{4p} + \frac{(b \, r + s/b)^2}{4p} \,.
\ee
Finally, using this result and  substituting the AGT map 
\be \label{N2AGT}
\al = a + Q/2 \,, \qquad \ep_1 = 1/b \,, \quad \ep_2 = b\,,
\ee
we find that when written with only simple zeros as explained above, the Kac determinant  (\ref{N2Kac}) becomes (up to an overall constant)
\be \label{N2KacInst}
\prod_{\substack{r\,s\le p \,n \\[1pt]  |r-s| = \pm \ell \! \! \! \! \mod \! p } } \!\!  \! \! \! \!   \! \!   (2a - r \ep_1 - s \ep_2)(2a + r \ep_1 + s \ep_2) \, \, = \!\!\!\!\! \prod_{\substack{r\,s\le p \,n \\[1pt]  |r-s| = \pm \ell \! \! \! \! \mod \! p } }   \! \! \! \!   \! \!   A_{r,s}\,.
\ee
In this expression it  is understood that only those factors for which $P_{\tilde{\ell}}(n-rs/p)$ is non-zero are included. 

It is now a simple matter to check that (\ref{N2KacInst}) agrees with the denominators in the instanton expansion. We should stress that this matching fixes the map (\ref{N2AGT})  which for general $p$ takes the standard AGT form. As an example, consider $p\!=\!7$ where the possible levels were given in  (\ref{p7levels}). Inserting this result into (\ref{N2KacInst}) we find agreement with the denominators in (\ref{Z27}). Note that from our previous result on the allowed levels we have $P_{3}(3/7)=P_{1}(5/7)=0$ implying e.g.~that there is no $A_{3,1}$ factor at level $n \!=\!6/7$, no $A_{2,2}$ factor at level $n\!=\!1$, and no $A_{5,1}$ factor at level  $n\!=\!10/7$. We have also checked the matching for many other terms and values of $p$ and have not found any discrepancies with the above conjecture.

Before continuing to the higher rank cases let us mention that there exists an intriguing  relation between the symmetry algebras of the $(N,p)\!=(2,p)$ cosets and the super $\cW$-algebra obtained by Drinfeld-Sokolov reduction from the $\widehat{D}(2|1;\al)$ super current algebra \cite{Semikhatov:2001}. It may be possible to use this connection to make additional checks of the proposed relation with gauge theory. Although perhaps it is only the ``bosonic version" of the symmetry algebra that is accessible in this approach. The bosonic version only contains generators with integer conformal dimensions (for instance the bosonic version of the $\cN\!=\!1$ superconformal algebra has generators of conformal dimension $2,4$ and $6$ and is discussed e.g.~in appendix E of \cite{Blumenhagen:1994}). 

Since the $N=p=2$ case is related to the super-Liouville theory one might naively expect that for some values of $p$ the higher rank cosets should be related to super-Toda theories. However, this seems unlikely since the central charges of the models that are conventionally called super-Toda theories  \cite{Evans:1990} do not match the coset central charge (in particular the super-Toda central charges are not symmetric under $b\leftrightarrow1/b$ except for the rank one case). 

The reason why the $N\!=\!p\!=\!2$ model is related to a superconformal algebra can be traced to the fact that when $p=N$ fields with half-integer conformal dimension appear in the coset theory (see e.g.~\cite[Section 7.3.2]{Bouwknegt:1992} and references therein). Therefore, it is natural to expect that e.g.~the $(N,p)=(3,3)$ model should correspond to a super-$\cW_3$ algebra. However, the obvious supersymmetrisation of $\cW_3$ is only consistent for $c=10/7$ \cite{Inami:1988}. This central charge agrees with the coset central charge (\ref{ccos}) when $\ka=1$ and indeed for this value the super-$\cW_3$ algebra arises from the coset \cite{Hornfeck:1990}.  Thus it seems that conformal blocks for the super-$\cW_3$ algebra should be related to instanton partition functions for the $\SU(3)$ gauge theory on $\RR/\ZZ_3$ with $\ep_{1,2}$ satisfying the constraint obtained by inserting $\ka=1$ into (\ref{kaep}). However, one puzzling aspect is that  instanton expansion has gaps which at least at first glance seems difficult to obtain from the CFT. The complicated symmetry algebra of the $(3,3)$ coset with $\ka\neq 1$ has been discussed in  \cite{Ahn:1990b}. 

In the general higher-rank case very little is known about the symmetry algebra, but the method based on the  Kac  determinant should be within reach. Consider first the $(N,1)$ models. For $N>3$ there have not been many explicit checks in the literature due to the complicated nature of the $\cW_N$ algebras. The  Kac determinant is known for the $\cW_N$ algebras for all $N$ \cite{Watts:1989} (see also  \cite[Section 6.4.1]{Bouwknegt:1992} and \cite{Kanno:2009}). When rewritten using the $A_{N-1}$ AGT map (here $\vec{\rho}$ is Weyl vector) 
\be \label{NAGT}
\vec{\al} = \vec{a} + (b{+}{\ts \frac{1}{b}})\vec{\rho}   \,, \qquad \ep_1 = 1/b \,, \quad \ep_2 = b\,,
\ee
and with all non-zero exponents set equal to 1 as above, the Kac determinant takes the form
\be \label{NKacInst}
\prod_{i<j} \prod_{r,s \le n }  (a_i-a_j - r \ep_1 - s \ep_2)(a_i-a_j + r \ep_1 + s \ep_2) \,  = \prod_{i
<j} \prod_{rs\le n }   \cA_{r,s}(i,j)\,.
\ee
We have checked for several cases that the denominators in the instanton expansion indeed agree with this expression.

When $p\!\ge\!1$ the Kac determinant has not been explicitly written down in the literature, but it should be possible to extract it from the results in \cite{Christe:1988} by using the method in \cite[Section 7.3.4]{Bouwknegt:1992}. The analogue of $P_{\tilde{\ell}}(n{-}rs/p)$ in (\ref{N2Kac}) --- the so called string functions --- are known, but in the general case their construction is a little involved, see \cite{Georgiev:1995}.  The allowed levels can be extracted from the string functions so it should also be possible to check if the gaps in the gauge theory instanton expansions are reproduced. 
By analogy with the $N\!=\!2$ case it seems likely that the Kac determinant takes a form similar to (\ref{NKacInst}); we have checked that the instanton expressions are consitent with this expectation. A complete analysis is left for future work.

\setcounter{equation}{0}
\section{Factorisation of coset conformal blocks   }\label{sFac}

Various relationships between conformal blocks for different coset theories have been studied in the literature, mainly for  rational theories, see e.g.~\cite{Douglas:1987}. 

Of particular relevance for us are the observations in \cite{Crnkovic:1989a} and  \cite{Lashkevich:1993}.  In these papers the rational versions of the coset (\ref{suncos}) were studied and conformal blocks for the $(N,p)$ model were argued to factorise into a certain product of $p$ conformal blocks for the $(N,1)$ model. This result originates in the following simple (formal) factorisation property of the coset
\be \label{faccos}
\cC_\ka \equiv \frac{ \widehat{\su}(N)_\ka \oplus \widehat{\su}(N)_p }{ \widehat{\su}(N)_{\ka+p} } 
 = \frac{\cF_{\ka} \cF_{\ka+1} \cdots \cF_{\ka+p-1}}{\cF_{1} \cF_{2} \cdots \cF_{p-1}}\,,
\ee
where we used the short-hand notation 
\be
\cF_\ka = \frac{ \widehat{\su}(N)_\ka \oplus \widehat{\su}(N)_1 }{ \widehat{\su}(N)_{\ka+1} } \,.
\ee
The factorisation property (\ref{faccos}) is more transparent if one uses the (group) $G{\times}G/G$ multiplicative notation for the coset rather than the (Lie algebra) $g{\oplus}g/g$ additive notation  that we have used so far.

A priori the factorisation (\ref{faccos}) is just a formal trick, but for minimal model cosets (where $\ka$ is an integer) the meaning of the product of cosets on right hand side of (\ref{faccos}) has been  made precise  \cite{Crnkovic:1989a,Lashkevich:1993}. Our goal is to try to extrapolate this result  to the non-rational setting where $\ka$ is a free parameter. 

We  first consider the special case $p\!=\!2$ and focus on the numerator in (\ref{faccos}).  
The relation between $\ka$ and the gauge theory $\ep_{1,2}$ parameters is  as in (\ref{kaep}) with $p\!=\!2$. We know from \cite{Alday:2009a} that there is a relation between the conformal blocks for $\cF_{\ka}$ and the $\SU(2)$ instanton partition function where the deformation parameters are related to $\ka$ via (\ref{kaep}) with $p\!=\!1$. This means that for each factor in the numerator in (\ref{faccos}) one can associate a corresponding set of  $\ep_{1,2}$ parameters. For clarity we denote the deformation parameters corresponding to the $(\ell{+}1)$th factor as $\vep^{(\ell)}_{1,2}$ where in the current example $\ell=0,1$.  The $\vep^{(\ell)}_{1,2}$ are not independent of the $\ep_{1,2}$ deformation parameters of the theory on $\RR^4/\ZZ_2$: By using the relation (\ref{kaep}) with $p\!=\!1$ and $\ka^{(0)}=\ka$ resp. $\ka^{(1)}=\ka\!+\!1$ we find the corresponding $\vep^{(\ell)}_{1,2}$ ($\ell=0,1$)
\bea \label{veps}
-2\, \frac{\ep_2}{\ep_1+\ep_2} = \ka{+}N  = - \, \frac{\vep^{(0)}_2}{\vep^{(0)}_1+\vep^{(0)}_2} \quad  \Rightarrow \qquad \frac{\vep^{(0)}_2}{\vep^{(0)}_1} =  \frac{\ep_2 {-} \ep_1}{2 \ep_1} \,, \non \\
-2\, \frac{\ep_2}{\ep_1+\ep_2} +1 = \ka{+}1{+}N  = - \, \frac{\vep^{(1)}_2}{\vep^{(1)}_1+\vep^{(1)}_2} \quad  \Rightarrow \qquad \frac{\vep^{(1)}_2}{\vep^{(1)}_1} =  \frac{2 \ep_2}{\ep_1{-}\ep_2}\,.
\eea
These relations suggest that the instanton partition function on $\RR/\ZZ_2$ should factorise into some sort of product of two $\RR^4$ instanton partition functions  with ``effective" deformation parameters as in (\ref{veps}).  

To understand in more detail what the product should be we need to recall some results from \cite{Crnkovic:1989a,Lashkevich:1993}. Initially we focus on the simplest case, $N\!=\!p\!=\!2$. The main idea in \cite{Crnkovic:1989a,Lashkevich:1993} is to look for the vertex operators (representations) of the tensor product of $\cC_\ka$ and $\cF_1$ in a certain projection of the vertex operators (representations) of the tensor product of $\cF_{\ka}$ and $\cF_{\ka+1}$. This implies that the conformal blocks of the $\cC_\ka$ theory are related to a projection of the product of the conformal blocks of $\cF_{\ka}$ and $\cF_{\ka+1}$. In particular, the internal states of the conformal blocks of $\cF_{\ka}$ and $\cF_{\ka+1}$ are not independent. We consider the simplest case of the irregular conformal block (as studied elsewhere in this paper). In the rational case (i.e.~for $\ka$ an integer) the conformal dimension for the simplest representation (called the vacuum sector in \cite{Crnkovic:1989a} and denoted $\ell=0$ in section \ref{sGen}) is given by
\be
 \De_{r,s}(p,\ka) = \frac{[(\ka+p+2)s-(\ka+2)r]^2-p^2}{4\,p\,(\ka+2)(\ka+p+2)}\,.
\ee
At the level of the conformal dimensions, the projection on the representations of the internal states in the conformal blocks is encoded in the relation \cite{Crnkovic:1989a}
\be \label{Deid}
\De_{r,s}(2,\ka) -\De_{r,v}(1,\ka) - \De_{v,s}(1,\ka{+}1) = -\half q^2,
\ee
where $q=v-(r+s)/2\in \ZZ$. The projected multiplication of conformal blocks of $\cF_\ka$ and $\cF_{\ka+1}$ is a sum of all products of conformal blocks whose internal states are consistent with (\ref{Deid}). The allowed possibilities are labelled by the integer $q$. 
To extrapolate to the non-rational case and to translate to gauge theory language we observe that (\ref{Deid}) can be written
\be \label{aid}
-\frac{a^2}{2} + \frac{(a + b\, q)^2}{2(1-b^2)} + \frac{(a + q/b)^2}{2(1-1/b^2)} = -\half q^2,
\ee
where
\be
a= \half( b\, r + s/b )\,, \qquad \qquad b^2= -\frac{\ka+2}{\ka+4}\,.
\ee
Using (\ref{veps}) the relation (\ref{aid}) can also be written 
\be \label{aid2}
 \frac{a^2}{2\ep_1\ep_2} - \sum_{\ell=0,1} \frac{(a + \ep^{(\ell)}_{\ell+1} \,q/2)^2}{\ep^{(\ell)}_1\ep^{(\ell)}_2}   =  \half q^2 \,.
\ee
Identifying the extrapolation of $a$ with the gauge theory Coulomb parameter we expect that the $\SU(2)$ instanton partition function on $\RR^4/\ZZ_2$ should be the sum over $q$ of the product of two $\RR^4$ instanton partition functions with Coulomb parameters $a + \ep^{(\ell)}_{\ell+1} \,q/2$.

This expectation is consistent with the known form of the gauge theory result \cite{Sasaki:2006,Gasparim:2008,Bruzzo:2009}:
\be \label{Zfac}
Z_2^{(2)} = \sum_{q\in \ZZ}  \frac{y^{\frac{ q^2}{2} }}{L_q(a,\ep_1,\ep_2) } Z_2(a+\frac{\vep_1^{(0)} q}{2},\vep^{(0)}_1,\vep^{(0)}_2,y)Z_2(a+\frac{\vep_2^{(1)} q}{2},\vep^{(1)}_1,\vep^{(1)}_2,y)\,,
\ee
where $Z_2$ is as in (2.1) and $\vec{a}=(a,-a)$. Note that the parameter $y$ used here is not the same as the one used elsewhere in this paper:  $y_{\rm here}=y^2_{\rm there}$.   

A few comments are in order. From the discussion in  \cite{Crnkovic:1989a,Lashkevich:1993}  we would also expect a contribution from the denominator $\cF_1$ (in addition to its role in the projection). However, we recall from the discussion in the introduction that the conjecture \cite{Nishioka:2011} is that the coset CFT (\ref{suncos}) tensored with the coset (\ref{paracos}) is what should be dual to the gauge theory. But we have
\be
\frac{ \widehat{\su}(p)_{N}  }{\hat{\ul}(1)^{p-1} }  = \cF_{1} \cF_{2} \cdots \cF_{p-1}\,, 
\ee
using well-known coset manipulations (see e.g.~\cite[Section 7.4.2]{Bouwknegt:1992}). Therefore any residual dependence on the denominator is likely to cancel against the prefactor. 
So far we have not discussed the quantities $L_q(a,\ep_1,\ep_2)$ that appear in the instanton partition function (\ref{Zfac}).  As recently discussed \cite{Bonelli:2011b} one way to interpret them is the following. If one also includes the perturbative contributions in both $Z_2$ and $Z_2^{(2)}$, i.e.~$Z\rar Z_{\rm full} = Z_{\rm pert}Z_{\rm inst}$ then the relation involving the $Z_{\rm full}$'s (the blow-up formula) essentially takes the same form as in (\ref{Zfac}) but without the $L_q(a,\ep_1,\ep_2)$ and $y^{q^2/2}$ pieces (see \cite{Bonelli:2011b} for further details). Based on AGT arguments it therefore seems  plausible to look for a CFT explanation of the $L_q(a,\ep_1,\ep_2)$ pieces starting from the theree-point functions.  Relations between the three-point functions were discussed in 
 \cite{Crnkovic:1989a,Lashkevich:1993} and it would be interesting to see if an extrapolation of these results can be used to explain the form of $L_q(a,\ep_1,\ep_2)$. Let us also note that in \cite{Bonelli:2011b} the identity (\ref{aid2}) was related to the matching of the classical pieces of the prepotential in the blow-up formula.

Clearly the above extrapolations of the rational CFT arguments in \cite{Crnkovic:1989a,Lashkevich:1993} are  heuristic and incomplete, but the results nevertheless strongly indicate that the factorisation property of the coset (\ref{faccos})  should provide the answer to the question raised in \cite{Bonelli:2011a,Bonelli:2011b} concerning the CFT interpretation of the factorised form of the gauge theory instanton partition functions on $\RR^4/\ZZ_p$.

The above discussion can also be extended to higher $p$. We close this section with some brief comments about these cases. 

For general $p$ we find for the $(\ell{+}1)$th factor of the coset ($\ell=0,1,\ldots,p-1$)
\be
-p\, \frac{\ep_2}{\ep_1+\ep_2}{+}\ell= \ka{+}N{+}\ell  = - \, \frac{\vep^{(\ell)}_2}{\vep^{(\ell)}_1+\vep^{(\ell)}_2} \;\;  \Rightarrow \quad \frac{\vep^{(\ell)}_2}{\vep^{(\ell)}_1} =  \frac{(\ell{+}1) \ep_2 -(p{-}\ell{-}1)\ep_1}{ (p{-}\ell)\ep_1 - \ell\ep_2}.
\ee
Again this result is consistent with the factorised instanton result since
\be \label{vepsp}
(\vep^{(\ell)}_1,\vep^{(\ell)}_2) = ( \, (p{-}\ell)\ep_1 {-}\ell \ep_2 \,,\, (\ell{+}1{-}p)\ep_1 + (\ell{+}1 ) \ep_2 \, ) \,,
\ee
are precisely the parameters corresponding to the action of $\U(1)^2$ on the the local coordinates of the dual cones of the blow up of $\RR^4/\ZZ_p=\CC^2/\ZZ_p$  (see e.g.~\cite{Fucito:2006,Nishioka:2011} for more details about the local coordinates). As usual,  $\U(1)^2$ acts on $z_{1,2}$ of $\CC^2$ as $z_{1,2}\rar e^{\ep_{1,2}}z_{1,2}$. 

The generalisation of the restrictions on the conformal dimensions is straightforward. For instance for $p=3$, $N=2$ we have (in the vacuum sector)
\be \label{Deid3}
\De_{r,s}(3,\ka) -\De_{r,v_1}(1,\ka) - \De_{v_1,v_2}(1,\ka{+}1)- \De_{v_2,s}(1,\ka{+}2) = -\half (q_1^2 + q_2^2-q_1 q_2) \,,
\ee
where 
\be
q_1 = v_1 - \frac{2}{3} r -  \frac{1}{3} s\,, \qquad q_2 = v_2 - \frac{1}{3} r -  \frac{2}{3} s\,.
\ee
Using arguments as above, the identity can be rewritten in the general form 
\be
 \frac{a^2}{p\,\ep_1\ep_2} - \sum_{\ell=0}^{p-1} \frac{(a + \ep^{(\ell)}_{1} q_{\ell+1}/2 + \ep^{(\ell)}_{2} q_{\ell}/2  )^2}{\ep^{(\ell)}_1\ep^{(\ell)}_2}   =  \frac{1}{4} q^\ell C_{\ell \tilde{\ell}} q^{\tilde{\ell}} \,,
\ee
where it is implicit that $q_0=q_p=0$ and $C_{\ell \tilde{\ell}} $ is the $A_{p-1}$ Cartan matrix. 
This result appears consistent with the instanton partition functions in \cite{Gasparim:2008}. We have only considered the vacuum sector. In this sector the $q_\ell$ are integers, but in other sectors this is likely no longer true, which should be kept in mind.  

\setcounter{equation}{0} 
\section{Discussion} \label{sdisc}

In this paper we studied the proposal \cite{Feigin:2011,Nishioka:2011} that $\SU(N)$ gauge theories on $\RR^4/\ZZ_p$ should be related to the para-Toda theories that are connected to the $(N,p)$ coset (\ref{suncos}). We found non-trivial evidence for the correctness of this proposal, but there are many things that need to be understood better and also many possible extensions. 

As noted  in section \ref{sS3}, one puzzling aspect of the relation between the (irregular) conformal blocks of the coset theory and the gauge theory instanton partition functions is that the gauge theory expressions do not seem to involve the CFT descendants at all levels nor all possible sectors of highest weight states. Conversely, on the gauge theory side there is an extension of the instanton partition function that involves $p{-}1$ additional variables (called $x_\ell$ in section \ref{sInst}) whose meaning on the CFT side is not yet clear. These facts possibly indicate that there is a more general version of the relation. 

Some obvious extensions are to include matter on the gauge theory side which corresponds to considering proper conformal blocks on the CFT side. This was accomplished for $(N,p)=(2,2)$ in \cite{Belavin:2011,Bonelli:2011b}. For the para-Liouville theories ($N\!=\!2$, general $p$) the three-point functions are known \cite{Bershtein:2010}. It would be interesting to compare these expressions to the perturbative parts of the $\SU(2)$ gauge theory on $\RR^4/\ZZ_p$ and in particular to recover the ``selection rules" noted in \cite{Bershtein:2010}. For higher rank and $p\!>\!1$, there are at present no CFT results, but the gauge theory results \cite{Gasparim:2008} indicate that for general $N$ and $p$ the three-point function (with restrictions as in \cite{Fateev:2007}) should factorise into $p$ pieces just like for $N\!=\!2$. 
It is also desirable to extend the analysis to the full CFT correlation functions which on the gauge theory side should correspond to extending the analysis of Pestun \cite{Pestun:2007} to gauge theories on $\RR^4/\ZZ_p$.

Another possible extension is to replace $\RR^4/\ZZ_p$ by a general toric singularty. It is known that in four (non-compact) dimensions the most general toric singularity is $\RR^4/\Ga_{p,q}$ where $p$, $q$ are coprime integers with $p>q>0$ (see e.g.~\cite{Fucito:2006} for a summary). In this language $\RR^4/\ZZ_p$ corresponds to $\RR^4/\Ga_{p,p-1}$. The dual CFT for the general $(p,q)$ case is not known at present, but the central charges can easily  be computed  from the anomaly polynomial of the $6d$ $A_{N-1}$ $(2,0)$ theory using the method in \cite{Benini:2009,Nishioka:2011} (assuming the method is applicable also for these cases). For instance, for the $\Ga_{p,1}$ case (also known as blown down $\cO_{\PP_1}(-p)$) it can be shown that 
\be
c^{p,1}_{\rm M-theory} =  2(N-1) + \frac{N^3{-}N}{p}\left[ \left(\sqrt{\frac{\ep_1}{\ep_2}} +\sqrt{\frac{\ep_2}{\ep_1}} \right)^2 -(p-2)^2\right] .
\ee
Note that this expression agrees with (\ref{cM}) when $p=2$ as it should. Another example is $\Ga_{5,2}$ where one finds 
\be
c^{5,2}_{\rm M-theory} =  3(N-1) +(N^3{-}N) \left[ \left(\sqrt{\frac{\ep_1}{\ep_2}} +\sqrt{\frac{\ep_2}{\ep_1}} \right)^2 -26\right] .
\ee
Another clue that might be helpful in identifying the dual CFT comes from the ``factorised" form of the instanton partition function \cite{Gasparim:2008,Bruzzo:2009} that should have a CFT explanation similar to the one in section \ref{sFac}. 

An appealing feature of the gauge theory results is that all theories are of roughly the same difficulty. In contrast, on the CFT side some cases appear more complicated that others (at least at first sight). This is in particular true for the theories whose symmetry algebra is non-abelianly braided. On the other hand, the gauge theory results indicate that also these theories should be amenable to analysis (at least for some sectors).

The simple Kac determinant method we proposed in section \ref{sGen} as a way to check AGT-type relations is presumably generally applicable. For instance, it could be used to investigate the relation  \cite{Braverman:2010}  between $\SU(N)$ gauge theories with a general surface operator and the corresponding general $\cW$-algebras that are labelled by partitions of $N$. The gauge theory instanton partition functions were determined in \cite{Wyllard:2010b}.  
There exists \cite{deBoer:1993} a construction algoritm for the general $\cW$-algebras but it is complicated and only some cases have been worked out. The approach based on the  Kac determinant offers a simpler alternative. So far the Kac determinant has only been explicitly written down for some cases (see e.g.~\cite{Kac:2004}) but more general cases should also be accessible.





\begingroup\raggedright\endgroup


\begin{thebibliography}{10}

\bibitem{Alday:2009a}
L.~F. Alday, D.~Gaiotto, and Y.~Tachikawa, ``{Liouville correlation functions
  from four-dimensional gauge theories},''
  {{\em Lett. Math. Phys.}
  {\bfseries 91} (2010) 167},
{{\ttfamily arXiv:0906.3219[hep-th]}}.

\bibitem{Wyllard:2009}
N.~Wyllard, ``{$A_{N-1}$ conformal Toda field theory correlation functions from
  conformal $\cN=2$ $\SU(N)$ quiver gauge theories},''
 {{\em JHEP} {\bfseries
  11} (2009) 002},
{{\ttfamily arXiv:0907.2189[hep-th]}}.

\bibitem{Feigin:2011}
V.~Belavin and B.~Feigin, ``{Super Liouville conformal blocks from $\cN=2$ SU(2)
  quiver gauge theories},''
 {{\em JHEP} {\bfseries 07}
  (2011) 079},
{{\ttfamily arXiv:1105.5800[hep-th]}}.

\bibitem{Bonelli:2011a}
G.~Bonelli, K.~Maruyoshi, and A.~Tanzini, ``{Instantons on ALE spaces and super
  Liouville conformal field theories},''
{{\ttfamily arXiv:1106.2505[hep-th]}}.

\bibitem{Belavin:2011}
A.~Belavin, V.~Belavin, and M.~Bershtein, ``{Instantons and $2d$ superconformal
  field theory},''{{\ttfamily
  arXiv:1106.4001[hep-th]}}.

\bibitem{Bonelli:2011b}
G.~Bonelli, K.~Maruyoshi, and A.~Tanzini, ``{Gauge theories on ALE space and
  super Liouville correlation functions},''
{{\ttfamily arXiv:1107.4609[hep-th]}}.

\bibitem{Nishioka:2011}
T.~Nishioka and Y.~Tachikawa, ``{Para-Liouville/Toda central charges from
  M5-branes},'' {{\ttfamily
  arXiv:1106.1172[hep-th]}}.

\bibitem{Goddard:1984}
P.~Goddard, A.~Kent, and D.~I. Olive, ``{Virasoro algebras and coset space
  models},'' {{\em
  Phys. Lett.} {\bfseries B152} (1985) 88}; \\
P.~Goddard, A.~Kent, and D.~I. Olive, ``{Unitary representations of the
  Virasoro and Supervirasoro algebras},''
  {{\em Commun. Math. Phys.}
  {\bfseries 103} (1986) 105--119}.

\bibitem{Gawedzki:1988a}
K.~Gawedzki and A.~Kupiainen, ``{$G/h$ conformal field theory from gauged WZW
  model},'' {{\em
  Phys. Lett.} {\bfseries B215} (1988) 119}; \\
  D.~Karabali, Q.-H. Park, H.~J. Schnitzer, and Z.~Yang, ``{A GKO construction
  based on a path integral formulation of gauged Wess-Zumino-Witten actions},''
  {{\em Phys. Lett.}
  {\bfseries B216} (1989) 307}; \\
K.~Gawedzki and A.~Kupiainen, ``{Coset construction from functional
  integrals},'' {{\em
  Nucl. Phys.} {\bfseries B320} (1989) 625};\\
D.~Karabali and H.~J. Schnitzer, ``{BRST quantization of the gauged WZW action
  and coset conformal field theories},''
  {{\em Nucl. Phys.}
  {\bfseries B329} (1990) 649}.

\bibitem{Bouwknegt:1992}
P.~Bouwknegt and K.~Schoutens, ``{$\cW$-symmetry in conformal field theory},''
  {{\em Phys. Rept.}
  {\bfseries 223} (1993) 183},
{{\ttfamily arXiv:hep-th/9210010}}.

\bibitem{Nemeschansky:1991}
D.~Nemeschansky and N.~P. Warner, ``{Topological matter, integrable models and
  fusion rings},'' {{\em
  Nucl. Phys.} {\bfseries B380} (1992) 241},
{{\ttfamily arXiv:hep-th/9110055}}; \\
A.~LeClair, D.~Nemeschansky, and N.~Warner, ``{S matrices for perturbed $\cN=2$
  superconformal field theory from quantum groups},''
  {{\em Nucl. Phys.}
  {\bfseries B390} (1993) 653},
  {{\ttfamily arXiv:hep-th/9206041}}.

\bibitem{Bershtein:2010}
M.~Bershtein, V.~Fateev, and A.~Litvinov, ``{Parafermionic polynomials, Selberg
  integrals and three-point correlation function in parafermionic Liouville
  field theory},''
 {{\em Nucl. Phys.}
  {\bfseries B847} (2011) 413},
 {{\ttfamily arXiv:1011.4090[hep-th]}}.

\bibitem{Benini:2009}
L.~F. Alday, F.~Benini, and Y.~Tachikawa, ``{Liouville/Toda central charges
  from M5-branes},''
  {{\em Phys. Rev.
  Lett.} {\bfseries 105} (2010) 141601},
{{\ttfamily arXiv:0909.4776[hep-th]}}.

\bibitem{Bonelli:2009}
G.~Bonelli and A.~Tanzini, ``{Hitchin systems, $\cN=2$ gauge theories and
  W-gravity},'' {{\em
  Phys. Lett.} {\bfseries B691} (2010) 111},
{{\ttfamily arXiv:0909.4031[hep-th]}}.

\bibitem{Moore:1997}
G.~W. Moore, N.~Nekrasov, and S.~Shatashvili, ``{Integrating over Higgs
  branches},'' {{\em Commun. Math.
  Phys.} {\bfseries 209} (2000) 97},
{{\ttfamily arXiv:hep-th/9712241}}; \\
A.~Losev, N.~Nekrasov, and S.~L. Shatashvili, ``{Testing Seiberg-Witten
  solution},''
{{\ttfamily arXiv:hep-th/9801061}}.

\bibitem{Nekrasov:2002}
N.~A. Nekrasov, ``{Seiberg-Witten prepotential from instanton counting},'' {\em
  Adv. Theor. Math. Phys.} {\bfseries 7} (2004) 831,
{{\ttfamily arXiv:hep-th/0206161}}.

\bibitem{Alday:2010}
L.~F. Alday and Y.~Tachikawa, ``{Affine SL(2) conformal blocks from $4d$ gauge
  theories},'' {{\em Lett.
  Math. Phys.} {\bfseries 94} (2010) 87},
{{\ttfamily arXiv:1005.4469[hep-th]}}; \\
C.~Koz\c{c}az, S.~Pasquetti, F.~Passerini, and N.~Wyllard, ``{Affine $\sll(N)$
  conformal blocks from $\cN=2$ $\SU(N)$ gauge theories},''
  {\em JHEP} {\bfseries 01} (2011) 045, 
{{\ttfamily arXiv:1008.1412[hep-th]}}.

\bibitem{Argyres:1991}
P.~C. Argyres, J.~M. Grochocinski, and S.~H. Tye, ``{Structure constants of the
  fractional supersymmetry chiral algebras},''
 {{\em Nucl.Phys.}
  {\bfseries B367} (1991) 217},
  {{\ttfamily arXiv:hep-th/9110052}}.

\bibitem{Fateev:1985b}
V.~A. Fateev and A.~B. Zamolodchikov, ``{Representations of the algebra of
  parafermion currents of spin 4/3 in two-dimensional conformal field theory.
  Minimal models and the tricritical $\ZZ_3$ Potts model},''
 {{\em
  Theor. Math. Phys.} {\bfseries 71} (1987) 451}.

\bibitem{Crnkovic:1989a}
C.~Crnkovi\'{c}, G.~Sotkov, and M.~Stanishkov, ``{Renormalization group flow
  for general $\SU(2)$ coset models},''
 {{\em Phys. Lett.}
  {\bfseries B226} (1989) 297}; \\
C.~Crnkovi\'{c}, R.~Paunov, G.~M. Sotkov, and M.~Stanishkov, ``{Fusions of
  conformal models},''
{{\em Nucl. Phys.}
  {\bfseries B336} (1990) 637}.

\bibitem{Lashkevich:1993}
M.~Y. Lashkevich, ``{Coset construction of minimal models},''
 {{\em Int. J. Mod. Phys.}
  {\bfseries A8} (1993) 5673},
  {{\ttfamily arXiv:hep-th/9304116}}.

\bibitem{Sasaki:2006}
T.~Sasaki, ``{$O(-2)$ blow-up formula via instanton calculus on
  $\widehat{\CC^2/\ZZ_2}$ and Weil conjecture},''
{{\ttfamily arXiv:hep-th/0603162}}.

\bibitem{Gasparim:2008}
E.~Gasparim and C.-C.~M. Liu, ``{The Nekrasov conjecture for toric surfaces},''
 {{\em Commun. Math. Phys.}
  {\bfseries 293} (2010) 661},
 {{\ttfamily arXiv:0808.0884[math.AG]}}.

\bibitem{Bruzzo:2009}
U.~Bruzzo, R.~Poghossian, and A.~Tanzini, ``{Poincare polynomial of moduli
  spaces of framed sheaves on (stacky) Hirzebruch surfaces},''
 {{\em Commun. Math. Phys.}
  {\bfseries 304} (2011) 395},
{{\ttfamily arXiv:0909.1458[math.AG]}}.

\bibitem{Fucito:2004b}
F.~Fucito, J.~F. Morales, and R.~Poghossian, ``{Multi instanton calculus on ALE
  spaces},'' {{\em
  Nucl. Phys.} {\bfseries B703} (2004) 518},
{{\ttfamily arXiv:hep-th/0406243}}.

\bibitem{Fujii:2005}
S.~{Fujii} and S.~{Minabe}, ``{A combinatorial study on quiver varieties},''
  {{\ttfamily
  arXiv:math/0510455}}.

\bibitem{Kronheimer:1990}
P.~B. Kronheimer and H.~Nakajima, ``{Yang-Mills instantons on ALE gravitational
  instantons},'' {\em Math. Ann.} {\bfseries 288} (1990) 263.

\bibitem{Fucito:2006}
F.~Fucito, J.~F. Morales, and R.~Poghossian, ``{Instanton on toric
  singularities and black hole countings},''
  {{\em JHEP} {\bfseries
  0612} (2006) 073}, {{\ttfamily
  arXiv:hep-th/0610154}}; \\
L.~Griguolo, D.~Seminara, R.~J. Szabo, and A.~Tanzini, ``{Black holes,
  instanton counting on toric singularities and $q$-deformed two-dimensional
  Yang-Mills theory},''
  {{\em Nucl. Phys.}
  {\bfseries B772} (2007) 1},
  {{\ttfamily arXiv:hep-th/0610155}}.

\bibitem{Dijkgraaf:2007}
R.~Dijkgraaf and P.~Su\l{}kowski, ``{Instantons on ALE spaces and orbifold
  partitions},'' {{\em
  JHEP} {\bfseries 03} (2008) 013},
{{\ttfamily arXiv:0712.1427[hep-th]}}.

\bibitem{Flume:2002}
R.~Flume and R.~Poghossian, ``{An algorithm for the microscopic evaluation of
  the coefficients of the Seiberg-Witten prepotential},''
 {{\em Int. J. Mod. Phys.}
  {\bfseries A18} (2003) 2541},
{{\ttfamily  arXiv:hep-th/0208176}}; \\
U.~Bruzzo, F.~Fucito, J.~F. Morales, and A.~Tanzini, ``{Multi-instanton
  calculus and equivariant cohomology},'' {\em JHEP} {\bfseries 05} (2003) 054,
{{\ttfamily  arXiv:hep-th/0211108}}; \\
H.~Nakajima and K.~Yoshioka, ``{Instanton counting on blowup. I},'' {\em
  Invent. Math.} {\bfseries 162} (2005) 313,
{{\ttfamily  arXiv:math/0306198}}; \\
N.~Nekrasov and A.~Okounkov, ``{Seiberg-Witten theory and random partitions},''
{{\ttfamily  arXiv:hep-th/0306238}}.

\bibitem{Argyres:1993}
P.~C. Argyres and S.~H. Tye, ``{Tree scattering amplitudes of the spin 4/3
  fractional superstring. 1. The untwisted sectors},''
  {{\em Phys.Rev.} {\bfseries
  D49} (1994) 5326},
 {{\ttfamily arXiv:hep-th/9310131}}.

\bibitem{Gaiotto:2009b}
D.~Gaiotto, ``{Asymptotically free $\cN=2$ theories and irregular conformal
  blocks},''
{{\ttfamily arXiv:0908.0307}}.

\bibitem{Marshakov:2009}
A.~Marshakov, A.~Mironov, and A.~Morozov, ``{On non-conformal limit of the AGT
  relations},'' {{\em
  Phys. Lett.} {\bfseries B682} (2009) 125},
{{\ttfamily arXiv:0909.2052[hep-th]}}.

\bibitem{Hadasz:2010}
L.~Hadasz, Z.~Jaskolski, and P.~Suchanek, ``{Proving the AGT relation for $N_f
  = 0,1,2$ antifundamentals},''
  {{\em JHEP} {\bfseries 06}
  (2010) 046},
{{\ttfamily arXiv:1004.1841[hep-th]}}.

\bibitem{Fateev:2009}
V.~A. Fateev and A.~V. Litvinov, ``{On AGT conjecture},''
 {{\em JHEP} {\bfseries 02}
  (2010) 014},
{{\ttfamily arXiv:0912.0504[hep-th]}}; \\
S.~{Yanagida}, ``{Norm of logarithmic primary of Virasoro algebra},''
 {{\em Lett. Math. Phys.}
  (2011) 51}, {{\ttfamily arXiv:1010.0528[math.QA]}}.

\bibitem{Kakushadze:1993}
Z.~Kakushadze and S.~H. Tye, ``{Kac and new determinants for fractional
  superconformal algebras},''
  {{\em Phys.Rev.} {\bfseries
  D49} (1994) 4122},
 {{\ttfamily arXiv:hep-th/9310160}}.

\bibitem{Kastor:1987}
D.~Kastor, E.~J. Martinec, and Z.-a. Qiu, ``{Current algebra and conformal
  discrete series},''
 {{\em Phys. Lett.}
  {\bfseries B200} (1988) 434}; \\
J.~Bagger, D.~Nemeschansky, and S.~Yankielowicz, ``{Virasoro algebras with
  central charge $c> 1$},''
  {{\em Phys. Rev. Lett.}
  {\bfseries 60} (1988) 389}; \\
F.~Ravanini, ``{An infinite class of new conformal field theories with extended
  algebras},'' {\em Mod. Phys. Lett.} {\bfseries 3A} (1988) 397.

\bibitem{Bagger:1988}
J.~Bagger and D.~Nemeschansky, ``{Coset constructions of chiral algebras},'' in
  {\em Strings '88 workshop}, S.~J. {Gates et.~al.}, ed., pp.~115--124.
\newblock World Scientific, 1989.

\bibitem{Semikhatov:2001}
A.~M. Semikhatov and B.~L. Feigin, ``{The $\widehat{\sll}(2)\oplus
  \widehat{\sll}(2)/\widehat{\sll}(2)$ coset theory as a Hamiltonian reduction
  of $\widehat{D}(2|1;\alpha)$ superalgebra},''
 {{\em JETP
  Lett.} {\bfseries 74} (2001) 59},
 {{\ttfamily arXiv:hep-th/0102078}}.

\bibitem{Blumenhagen:1994}
R.~Blumenhagen, W.~Eholzer, A.~Honecker, K.~Hornfeck, and R.~H{\"{u}}bel,
  ``{Coset realization of unifying $\cW$-algebras},''
 {{\em Int. J. Mod. Phys.}
  {\bfseries A10} (1995) 2367},
  {{\ttfamily arXiv:hep-th/9406203}}.

\bibitem{Evans:1990}
J.~Evans and T.~J. Hollowood, ``{Supersymmetric Toda field theories},''
  {{\em Nucl. Phys.} {\bfseries B352} (1991)
  723}.

\bibitem{Inami:1988}
T.~Inami, Y.~Matsuo, and I.~Yamanaka, ``{Extended conformal algebras with
  $\cN=1$ supersymmetry},''
 {{\em Phys. Lett.}
  {\bfseries B215} (1988) 701}.

\bibitem{Hornfeck:1990}
K.~Hornfeck and E.~Ragoucy, ``{A coset construction for the super
  $\cW_3$-algebra},''
 {{\em Nucl. Phys.}
  {\bfseries B340} (1990) 225}.

\bibitem{Ahn:1990b}
C.-h. Ahn, K.~Schoutens, and A.~Sevrin, ``{The full structure of the super
  $\cW_3$ algebra},'' {{\em
  Int. J. Mod. Phys.} {\bfseries A6} (1991) 3467}.

\bibitem{Watts:1989}
G.~M.~T. Watts, ``{Determinant formul\ae{} for extended algebras in
  two-dimensional conformal field theory},''
  {{\em Nucl. Phys.}
  {\bfseries B326} (1989) 648}.

\bibitem{Kanno:2009}
S.~Kanno, Y.~Matsuo, S.~Shiba, and Y.~Tachikawa, ``{$\cN\!=\!2$ gauge theories
  and degenerate fields of Toda theory},''
 {{\em Phys. Rev.}
  {\bfseries D81} (2010) 046004},
{{\ttfamily arXiv:0911.4787[hep-th]}}.

\bibitem{Christe:1988}
P.~Christe and F.~Ravanini, ``{$G_N \otimes G_L / G_{N+L}$ conformal field
  theories and their modular invariant partition functions},''
{{\em Int. J. Mod. Phys.}
  {\bfseries A4} (1989) 897}; \\
J.~Soda and H.~Yoshii, ``{Kac formulas for the extended Virasoro algebras},''
 {{\em Prog. Theor. Phys.} {\bfseries
  80} (1988) 941}; \\
S.~Nam, ``{Perturbed $c>1$ conformal field theories and generalized Toda field
  theories},'' {{\em
  Phys. Lett.} {\bfseries B243} (1990) 231}.

\bibitem{Georgiev:1995}
G.~Georgiev, ``{Combinatorial constructions of modules for infinite-dimensional
  Lie algebras, II. Parafermionic space},''
  {{\ttfamily arXiv:q-alg/9504024}}; \\
E.~{Ardonne}, R.~{Kedem}, and M.~{Stone}, ``{Fermionic characters and arbitrary
  highest-weight integrable $\sll_{r+1}$-modules},''
 {{\em Commun. Math. Phys.}
  {\bfseries 264} (2006) 427},
 {{\ttfamily
  arXiv:math/0504364[math.RT]}}; \\
E.~{Ardonne}, R.~{Kedem}, and M.~{Stone}, ``{Fusion products, Kostka
  polynomials and fermionic characters of $\widehat{\su}(r{+}1)_k$},''
 {{\em J. Phys.}
  {\bfseries A38} (2005) 9183},
 {{\ttfamily
  arXiv:math-ph/0506071}}.

\bibitem{Douglas:1987}
M.~R. Douglas, ``{$G/H$ conformal field theory}.'' CALT-68-1453, 1987; \\
M.~B. Halpern and N.~A. Obers, ``{Ward identities for affine Virasoro
  correlators},'' {{\em
  Int. J. Mod. Phys.} {\bfseries A9} (1994) 265},
 {{\ttfamily arXiv:hep-th/9207071}}; \\
M.~Y. Lashkevich, ``{Conformal blocks of coset construction: Zero ghost
  number},'' {{\ttfamily
  arXiv:hep-th/9301094}}.

\bibitem{Fateev:2007}
V.~A. Fateev and A.~V. Litvinov, ``{Correlation functions in conformal Toda
  field theory I},''
 {{\em JHEP} {\bfseries
  11} (2007) 002},
{{\ttfamily arXiv:0709.3806[hep-th]}}.

\bibitem{Pestun:2007}
V.~Pestun, ``{Localization of gauge theory on a four-sphere and supersymmetric
  Wilson loops},''
{{\ttfamily arXiv:0712.2824[hep-th]}}.

\bibitem{Braverman:2010}
A.~Braverman, B.~Feigin, L.~Rybnikov, and M.~Finkelberg, ``{A finite analog of
  the AGT relation I: finite $\cW$-algebras and quasimaps' spaces},''
{{\ttfamily arXiv:1008.3655[math.AG]}}; \\
N.~Wyllard, ``{$\cW$-algebras and surface operators in $\cN=2$ gauge
  theories},'' {{\em J.~Phys.} {\bfseries A44} (2011) 155401},
  {{\ttfamily arXiv:1011.0289[hep-th]}}.

\bibitem{Wyllard:2010b}
N.~Wyllard, ``{Instanton partition functions in $\cN=2$ $\SU(N)$ gauge theories
  with a general surface operator, and their $\cW$-algebra duals},''
  {{\em JHEP} {\bfseries 1102}
  (2011) 114}, {{\ttfamily arXiv:1012.1355[hep-th]}}; \\
H.~Kanno and Y.~Tachikawa, ``{Instanton counting with a surface operator and
  the chain-saw quiver},''
  {{\em JHEP} {\bfseries 1106}
  (2011) 119}, {{\ttfamily arXiv:1105.0357[hep-th]}}.

\bibitem{deBoer:1993}
J.~de~Boer and T.~Tjin, ``{The relation between quantum $\cW$ algebras and Lie
  algebras},'' {{\em Commun. Math.
  Phys.} {\bfseries 160} (1994) 317},
{{\ttfamily arXiv:hep-th/9302006}}.

\bibitem{Kac:2004}
V.~G. {Kac} and M.~{Wakimoto}, ``{Quantum reduction in the twisted case},''
  {\em Progress in Math.} {\bfseries 237} (2005) 85,
 {{\ttfamily arXiv:math-ph/0404049}}; \\
B.~Noyvert, ``{Ramond sector of superconformal algebras via quantum
  reduction},'' {\em JHEP} {\bfseries 11} (2006) 045,
{{\ttfamily arXiv:math-ph/0408061}}.

\end{thebibliography}
\end{document}